\documentclass[aps,pra,twocolumn,superscriptaddress]{revtex4-2} 

\usepackage[final,authormarkup=none,defaultcolor=purple]{changes}
\definechangesauthor[name={Benedikt}, color=purple]{BT}
\definechangesauthor[name={Anders}, color=purple]{AS}
\definechangesauthor[name={Emil}, color=purple]{EH}
\definechangesauthor[name={Soubhadra}, color=purple]{SM}

\usepackage[]{graphicx}
\usepackage[dvipsnames]{xcolor}
\usepackage[inkscapelatex=false]{svg}
\usepackage[colorlinks=true,citecolor=NavyBlue,linkcolor=RubineRed,urlcolor=Cerulean,pdfencoding=auto]{hyperref}
\usepackage[]{amsmath}
\usepackage[]{amssymb}
\def\enquote#1{``#1''}
\def\bra#1{\mathinner{\langle{#1}|}}
\def\ket#1{\mathinner{|{#1}\rangle}}
\def\braket#1{\mathinner{\langle{#1}\rangle}}

\def\abs#1{\mathinner{|{#1}|}}
\def\abssq#1{\mathinner{|{#1}|}^2}
\def\proj#1{\ket{#1}\bra{#1}}
\def\op#1#2{\ket{#1}\bra{#2}}

\DeclareMathOperator{\tr}{tr}

\begin{document}
\title{Hybrid Single-Ion Atomic-Ensemble Node for High-Rate Remote Entanglement Generation}

\author{Benedikt Tissot}
\email[]{benedikt.tissot@nbi.ku.dk}
\affiliation{Center for Hybrid Quantum Networks (Hy-Q), Niels Bohr Institute, University of Copenhagen, Jagtvej 155A, DK-2200 Copenhagen, Denmark}

\author{Soubhadra Maiti}
\affiliation{QuTech, Delft University of Technology, Lorentzweg 1, 2628 CJ Delft, The Netherlands}
\affiliation{EEMCS, Quantum Computer Science, Delft University of Technology,
Mekelweg 4, 2628 CD Delft, The Netherlands}

\author{Emil R. Hellebek}
\affiliation{Center for Hybrid Quantum Networks (Hy-Q), Niels Bohr Institute, University of Copenhagen, Jagtvej 155A, DK-2200 Copenhagen, Denmark}

\author{Anders Søndberg Sørensen}
\email[]{anders.sorensen@nbi.ku.dk}
\affiliation{Center for Hybrid Quantum Networks (Hy-Q), Niels Bohr Institute, University of Copenhagen, Jagtvej 155A, DK-2200 Copenhagen, Denmark}

\begin{abstract}
  Different quantum systems possess different favorable qualities.
  On the one hand, ensemble-based quantum memories are suited for fast multiplexed long-range entanglement generation.
  On the other hand, single-atomic systems provide access to gates for processing of information.
  Both of those can provide advantages for high-rate entanglement generation within quantum networks.
  We develop a hybrid architecture that takes advantage of these properties by combining trapped-ion nodes and nodes comprised of spontaneous parametric down conversion photon pair sources and absorptive memories based on rare-earth ion ensembles.
  To this end, we solve the central challenge of matching the different bandwidths of photons emitted by those systems in an initial entanglement-generation step. 
  This enables the parallel execution of multiple probabilistic tasks in the initial stage.
  \replaced[id=BT]{As a particular example, w}{W}e show that our approach can lead to a significant speed-up for the fundamental task of creating ion-ion entanglement over hundreds of kilometers in a quantum network.
\end{abstract}
\maketitle

\added[id=BT]{\emph{Introduction}---}Long-range entanglement generation is central to building a quantum internet \cite{kimble-2008-quant_inter,wehner-2018-quant_inter}.
Therein, the entanglement can be used for secure communication \cite{ekert-1991-quant_crypt_based_bells_theor,acin-2007-devic_indep_secur_quant_crypt,gisin-2007-quant_commun,pirandola-2020-advan_quant_crypt}, enhanced sensing \cite{wasilewski-2010-quant_noise_limit_entan_assis_magnet,cassens-2025-entan_enhan_atomic_gravim}, distributed computing \cite{cuomo-2020-towar_distr_quant_comput_ecosy}, as well as fundamental physics experiments like the violation of Bell-inequalities \cite{bell-1964-einst_podol_rosen_parad,clauser-1969-propos_exper_to_test_local,hensen-2015-looph_free_bell_inequal_violat,giustina-2015-signif_looph_free_test_bells,storz-2023-looph_free_bell_inequal_violat}.
While entanglement generation between separate matter systems has been demonstrated in various physical implementations -- e.g., between atomic ensembles \cite{chou-2007-funct_quant_nodes_entan_distr,yuan-2008-exper_demon_bdcz_quant_repeat_node,yu-2020-entan_two_quant_memor_via}, trapped ions   \cite{moehring-2007-entan_singl_atom_quant_bits_at_distan,stephenson-2020-high_rate_high_fidel_entan,krutyanskiy-2023-entan_trapp_ion_qubit_separ_by_meter}, color centers in solids \cite{bernien-2013-heral_entan_between_solid_state,sipahigil-2016-integ_diamon_nanop_platf_quant_optic_networ,humphreys-2018-deter_deliv_remot_entan_quant_networ}, and rare earth ions \cite{lago-rivera-2021-telec_heral_entan_between_multim,liu-2021-heral_entan_distr_between_two,ruskuc-2025-multip_entan_multi_emitt_quant_networ_nodes} --
high-rate entanglement generation over hundreds of kilometers remains a challenge.

Trapped ions are a favorable candidate for quantum network nodes,
because they are established computational qubits featuring high fidelity quantum gates \cite{bruzewicz-2019-trapp_ion_quant_comput,moses-2023-race_track_trapp_ion_quant_proces} and long coherence times \cite{wang-2021-singl_ion_qubit_with_estim}.
Local one- and two-qubit gates allow implementing network protocols, e.g.,
repeaters, dividing long-distance entanglement into smaller segments to overcome exponential loss in optical fibers \cite{riebe-2008-deter_entan_swapp_with_ion,sangouard-2011-quant_repeat_based_atomic_ensem_linear_optic,beukers-2024-remot_entan_protoc_station_qubit}.
An outstanding problem, however, is that the entanglement generation rate is slow over long distances.

To achieve a faster rate, one can employ multiplexing, which boosts the rate of possible entanglement generation attempts of a single qubit from \(c/L\) over a distance \(L\) with \(c\) being  the speed of light 
to \(N c / L\) for \(N\) qubits \cite{simon-2007-quant_repeat_with_photon_pair,ruskuc-2025-multip_entan_multi_emitt_quant_networ_nodes}.
Nodes combining photon pair sources based on spontaneous parametric down conversion and an absorptive memory using a rare-earth ion doped crystal (SPDC+M)
have  demonstrated high  multiplexing capabilities and feature high duty cycles \cite{simon-2007-quant_repeat_with_photon_pair,sinclair-2014-spect_multip_scalab_quant_photon,lago-rivera-2021-telec_heral_entan_between_multim,liu-2021-heral_entan_distr_between_two,ruskuc-2025-multip_entan_multi_emitt_quant_networ_nodes} and long storage durations \cite{ruskuc-2022-nuclear_spin_wave_quant_regis,zhong-2015-optic_addres_nuclear_spins_solid}.
Single SPDC+M nodes have been shown to support more than \(1000\) modes \cite{businger-2022-non_class_correl_over_modes},
which is more feasible and cost-effective than trying to link \(1000\) trapped ions in parallel.
Although the high number of memory modes makes ensemble nodes ideal for entanglement generation over an elementary link,
due to the absence of deterministic gates, information \replaced[id=AS]{should preferably}{has to} be transferred to other systems for processing.
Repeaters can be implemented using linear optics \cite{sangouard-2011-quant_repeat_based_atomic_ensem_linear_optic}, but in this case, the probabilistic nature of the entanglement swapping (henceforth swap for brevity) limits the entanglement generation rate for long distances \cite{simon-2007-quant_repeat_with_photon_pair}.

In this letter, we propose a hybrid architecture tailored towards high-rate long-distance entanglement generation by uniting an SPDC+M backbone (BB) and trapped-ion edge-nodes (EN).
Both systems can be entangled with photons with high fidelity \cite{clausen-2011-quant_storag_photon_entan_cryst,saglamyurek-2011-broad_waveg_quant_memor_entan_photon,bock-2018-high_fidel_entan_between_trapp,saha-2025-high_fidel_remot_entan_trapp},
which can interface the systems. 
However, this is challenged by the mismatch of the intrinsically \emph{narrow} bandwidth of photons emitted by ions, and the \emph{broad-band} nature of SPDC photons.
One approach to overcome this is photon (re-)shaping \cite{vasilev-2010-singl_photon_made_to_measur,farrera-2016-gener_singl_photon_with_highl,morin-2019-deter_shapin_reshap_singl_photon,tissot-2024-effic_high_fidel_flyin_qubit_shapin} within a final swap between the ion and the memory \deleted[id=BT]{[changed citations]}\cite{cussenot-2025-unitin_quant_proces_nodes_cavit,sun-2025-hybrid_quant_repeat_chain_with}.
\added[id=BT]{These approaches rely on reshaping the stored photon, which can introduce additional losses as well as decoherence of the stored photon-photon entanglement.}
\added[id=BT]{Another option is to replace the SPDC with a photon pair emitter compatible with the atomic system, e.g., Rb atoms  both as the atomic system and for photon generation \cite{gu-2025-hybrid_quant_repeat_with_ensem}.}

\begin{figure*}[ht]
\centering
\includegraphics[width=\linewidth]{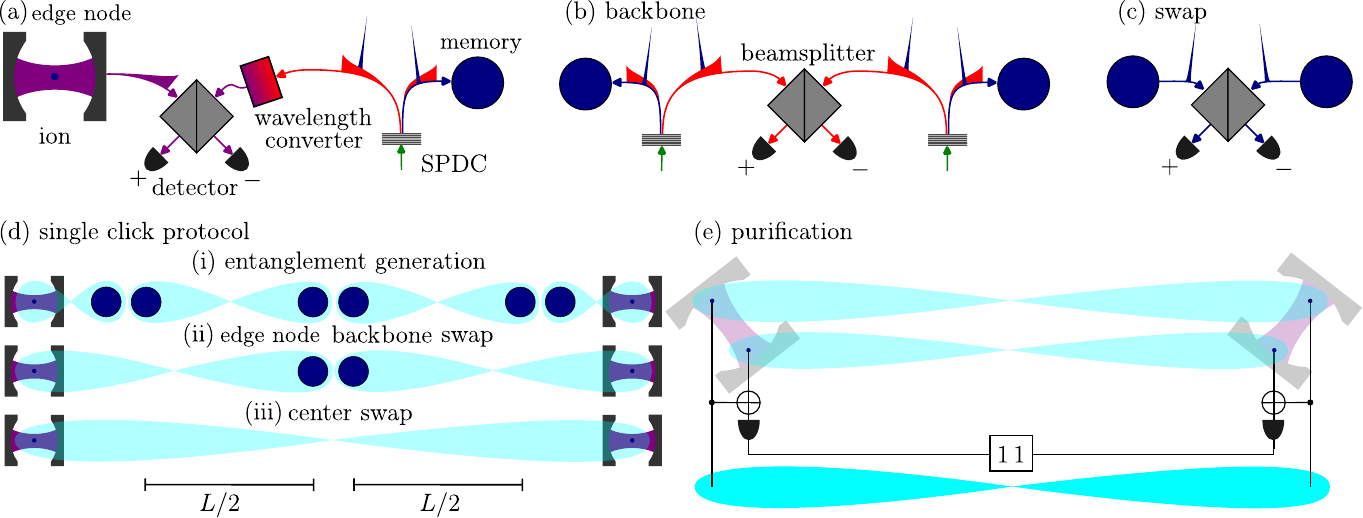}
\caption{\label{fig:atom-atom-full}Sketch of the protocol and \replaced[id=AS]{extension to long-distance entanglement generation}{device setups}. (a) \replaced[id=AS]{The proposed scheme to bridge the bandwidth difference of trapped ions and ensemble based memories (M).}{The edge-nodes (EN) consist of a trapped ion (in a cavity) and a}\added[id=BT]{A} spontaneous parametric down conversion source \replaced[id=BT]{ (SPDC) bridges the two systems.}{with attached multi-mode memory (SPDC+M) connected by fiber optics.}
After wavelength converting the SPDC photon to match the ion wavelength, the photons are interfered on a beamsplitter (BS) and measured with heralding detectors labeled \(\pm\).
\added[id=BT]{The hybrid nodes can then be embeded in an ensemble based quantum network as edge-nodes (EN) to extend the protocol to long-distance entanglement generation.}
(b) 
The \added[id=BT]{long-distance }backbone is composed entirely of SPDC+M nodes interconnected by optical fibers and heralding detectors.
(c) Memory-memory swaps are also implemented using fiber optics, BS, and heralding detectors. 
(d) Ion-ion entanglement \deleted[id=BT]{generation }is generated using a single-click protocol over the long distance \(L\) spanned by the backbone network. To this end, (i) backbone and edge-node entanglement is generated in parallel using the setups depicted in (a),(b); (ii) a swap (c) is performed between the edge-nodes and the closest backbone link. (iii) Finally, a swap in the center links the ions. (e) Performing the single-click scheme twice enables the use of purification. We consider a purification where local CNOTs followed by measurement of both controlled systems in $\ket{1}$ post-selects a Bell state of higher fidelity.}
\end{figure*}

We propose a novel approach
where we instead use the multi-mode nature of SPDC photon pairs by matching the photon correlation time of the SPDC to the broad bandwidth of the quantum memory, while \added[id=BT]{the} SPDC is modulated proportional to the slowly varying
temporal mode of the photon emitted by the ion.
Thereby we overcome the orders of magnitude different emission time-scales (SPDC \(\sim 100\,\)ns \cite{lago-rivera-2021-telec_heral_entan_between_multim} and trapped ions in a cavity \(\sim10\,\){\textmu}s \cite{meraner-2020-indis_photon_from_trapp_ion,krutyanskiy-2023-entan_trapp_ion_qubit_separ_by_meter}).
Furthermore, our approach
exploits short-range connections for the (likely) most inefficient links between the ions and the ensembles.
As a result, these links can be generated at a high repetition rate to compensate for the limited probability.\deleted[id=BT]{.}
The long-range entanglement\added[id=AS]{ generation}, which has to be performed at a slower rate due to the long communication time, is on the other hand enhanced by ensemble multiplexing.
Performing the matching and long-range generation simultaneously furthermore avoids unnecessary idle time.
With this method, rapid long-range entanglement can be established in a full network of links between different trapped ions\added[id=SM]{, see Fig.~\ref{fig:atom-atom-full}}. 
Local deterministic entanglement swaps between the ions can then further extend the range and complexity of the network.
\added[id=BT]{In a companion paper} \cite{ART} \added[id=BT]{we provide additional theoretical details on modeling the network and compare a single and double click protocol, while we focus on the idea of the SPDC matching in this letter and only use the protocols of Ref.~\cite{ART} to highlight the feasibility of the approach.}

\deleted[id=BT]{
  We focus on long-distance entanglement generation and thus consider a single-click scheme, but note that for shorter links or efficient ensemble memories, two-click \mbox{ \cite{barrett-2005-effic_high_fidel_quant_comput,sangouard-2011-quant_repeat_based_atomic_ensem_linear_optic,beukers-2024-remot_entan_protoc_station_qubit} } or reflection-based \mbox{ \cite{nemoto-2014-photon_archit_scalab_quant_infor_proces_diamon,omlor-2025-entan_gener_using_singl_photon} } protocols might be beneficial.
The entanglement generation protocol is illustrated in Fig.~\ref{fig:atom-atom-full}(d) and (e), and can be summarized in four steps.
It proceeds by (i) simultaneously generating entanglement links using single-click heralding between ions and memories  locally at ENs [Fig.~\ref{fig:atom-atom-full}(a)]
and between BB memories over a long distance \(L\) [Fig.~\ref{fig:atom-atom-full}(b)].
(ii) As soon as neighboring BB and EN links are ready, entanglement is swapped between them by heralding on a single click after releasing the photons from the memories [Fig.~\ref{fig:atom-atom-full}(c)].
Finally (iii) another optical swap between memories 
at the central repeater, connects the two halves.
}

\deleted[id=BT]{
We include a repeater in the center of the fundamental link, because parallelization
allows the additional swap to extend the distance, without significantly decreasing the success rate \mbox{ \footnote{
Further probabilistic repeaters would not benefit in the same way from the parallelization, which is why we propose to only use the probabilistic swaps in \enquote{fundamental} links and to extend range and complexity of the network beyond the fundamental link using deterministic ion swaps.
}}.
Alternatively, without the central repeater, a final swap can be performed between the BB and the remaining EN.
Upon successfully heralding every step, an ion-ion entangled link is achieved.
}

\deleted[id=BT]{
The \emph{last part} (iv) is purification to distill two ion-ion links into a single link with enhanced fidelity, see Fig.~\ref{fig:atom-atom-full}(e).
It proceeds by preparing two ion-ion links using the steps above and performing a CNOT gate between the local ions.
Measurement of both target ions in state \(\ket{1}\) heralds the purification of the ion-ion entanglement, see Fig.~\ref{fig:atom-atom-full}(e).
This post-selection helps in suppressing events where  the memories did not store any photons while retaining the better scaling of single-click protocols with the (long-distance) fiber losses \mbox{ \cite{duan-2001-long_distan_quant_commun_with,simon-2007-quant_repeat_with_photon_pair,zhao-2007-robus_creat_entan_between_remot_memor_qubit,sangouard-2011-quant_repeat_based_atomic_ensem_linear_optic,beukers-2024-remot_entan_protoc_station_qubit} }.
Although the single-click protocol inherently relies on phase stability \mbox{ \cite{sangouard-2011-quant_repeat_based_atomic_ensem_linear_optic,beukers-2024-remot_entan_protoc_station_qubit,ART} },
the purification  can reduce the sensitivity to such fluctuations if the phase is stable between two successful entanglement generation events \mbox{ \cite{ART}  }.  
}

\begin{figure}
\centering
\includegraphics[width=\linewidth]{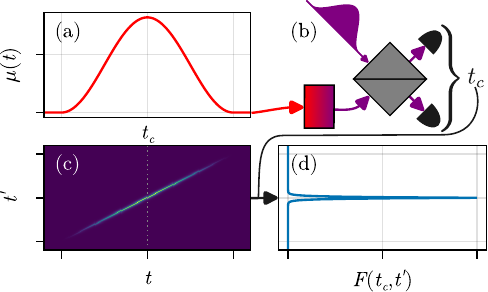}
\caption{\label{fig:SPDC-correlation}Matching the SPDC and ion\replaced[id=AS]{-}{ }photon flux within the edge-nodes. The correlation function of the photon pair \(\mu(t)F(t,t')\) of all time-bins 
  (c) emitted by the SPDC is split into a slowly varying envelope $\mu(t)$ 
  (a) that we can use to match to the atomic photon $\nu(t)$ (b).
  \added[id=BT]{Note that (a) and (c) display the combination of all time-bins.}
  Conditioned on the heralding click, (d) a short photon is stored within 
  one of the time-bins of 
  the multi-mode (ensemble) memory unless it is lost.
  For illustration we use \(T_a / T_c = 100\).
}
\end{figure}

\added[id=BT]{\emph{SPDC-matching}---} \added[id=AS]{The main step in interfacing the two systems is to generate entanglement between the trapped ion and the BB memory using the setup depicted in Fig.~\ref{fig:atom-atom-full}(a) and Fig.~\ref{fig:SPDC-correlation}.} 
\replaced[id=BT]{This is achived by the}{A} central \replaced[id=BT]{idea}{part} of our proposal\replaced[id=BT]{:}{ is} a method to bridge the different intrinsic time scales of the systems \added[id=BT]{by modulating an SPDC source}.
To describe this, we model the matter systems using the combined photonic and matter state immediately after the emission.
To implement a single-click protocol \cite{beukers-2024-remot_entan_protoc_station_qubit},
we consider the \emph{ions} to emit photons conditioned on the \replaced[id=BT]{matter}{internal} state,
\begin{align}
  \label{eq:ini-atom}
  \ket{\Psi_a} = \left[ \alpha_0 \ket{0} + \alpha_1 \ket{1} \int_{\mathbb{R}} dt \nu(t) a^{\dag}(t) \right] \ket{\emptyset_a},
\end{align}
with the matter states \(\ket{0}\) and \(\ket{1}\), the emission amplitude \(\alpha_1\) (\(\alpha_0 = \sqrt{1-\abssq{\alpha_1}}\)), the photonic creation operator \(a^{\dag}\) with the channel vacuum state \(\ket{\emptyset_a}\), and the normalized temporal mode function \(\nu(t)\) of duration $T_a$.
A state of this form can, e.g., be generated using stimulated Raman emission
\cite{keller-2004-contin_gener_singl_photon_with,barros-2009-deter_singl_photon_sourc_from_singl_ion,tissot-2024-effic_high_fidel_flyin_qubit_shapin}\added[id=BT]{ which was recently used for ion-ion entanglement generation \cite{liu-2026-long_lived_remot_ion_ion}}.

For (quasi) continuous driving of an SPDC\added[id=BT]{ source,} successful events are typically defined by the detection of a photon in one arm in a time window \replaced[id=AS]{defined by}{around} the detection time in the other arm.
\replaced[id=BT]{For convenient modeling}{To describe this} we divide the time interval into time\replaced[id=BT]{-}{ }bins \replaced[id=BT]{(which is not strictly necessary, but can be understood as the state in}{corresponding to} the acceptance interval\added[id=BT]{)}.
We take the SPDC+M state within a time-bin as the multi-modal state
\begin{align}
  \label{eq:ini-spdc}
& \ket{\Psi_b}  = \Big[ \beta_0 + \beta_1 \int_{\mathbb{R}^2} dt dt' \mu(t) F(t,t') b^{\dag}(t) c^{\dag}(t') \\
& \ + \beta_2 \int_{\mathbb{R}^4} d\vec{t} \mu(t_1) \mu(t_2) G(\vec{t}) b^{\dag}(t_1) b^{\dag}(t_2) c^{\dag}(t_3) c^{\dag}(t_4) \Big] \ket{\emptyset} , \notag
\end{align}
based on the detailed model developed in Ref.~\cite{hellebek-2024-charac_multim_natur_singl_photon}.
The operator \(b^{\dag}\) (\(c^{\dag}\)) corresponds to creating a photon in the channel connected with the BS and detectors (the memory), see Fig.~\ref{fig:atom-atom-full}.
Here \(\beta_1\) (\(\beta_2\)) denote the single (two) photon emission amplitude within a time-bin.
The vacuum state \(\ket{\emptyset}\) has the amplitude \(\beta_0\).
The temporal mode $\mu$ captures the overall shape of the first detected photon (in the $b$ mode), while the multi-modal and multi-photon character is contained in the multi-time-dependence of the \replaced[id=AS]{[}{(}double\replaced[id=AS]{]}{)} pair correlation function $F(t,t')$ [$G(\vec{t})$].
The correlation time $T_c$ between the photons of a pair corresponds to the width of $F(t,t')$ in the time difference $t'-t$. 
As the aim is to store most of the photon while avoiding storage of uncorrelated photons, we choose the acceptance interval and therefore the time-bin to be a few times the correlation time of \(F(t,t')\) (see Fig.~\ref{fig:SPDC-correlation}) but shorter than the much longer ion\replaced[id=AS]{-}{ }photon duration.

For the optical entanglement generation, as well as the probabilistic swaps \added[id=BT]{used for embedding the hybrid nodes in a full network}, we consider a setup that combines two input channels using a 50:50 beamsplitter with the two beamsplitter outputs terminating in detectors, see Fig.~\ref{fig:atom-atom-full}(a)--(c).
For both entanglement generation and swapping we use a single click to herald success, and
we model the optics for channels \(a\) and \(b\) using
\(a^{\dag} \to \sqrt{{\eta}/{2}}  ( d_+^{\dag} + d_-^{\dag} ) + \sqrt{1 - \eta} {a}_L^{\dag}\)
and
  \(b^{\dag} \to \sqrt{{\eta'}/{2}} ( d_+^{\dag} - d_-^{\dag} ) + \sqrt{1 - \eta'} {b}_L^{\dag}\),
see Ref.~\cite{ART} for additional details.
Here, the combined efficiencies \(\eta\) and \(\eta'\) account for all losses from emission to the detection of a click and can account for differences in the branches before the beamsplitter, e.g., frequency conversion \cite{meraner-2020-indis_photon_from_trapp_ion,krutyanskiy-2024-multim_ion_photon_entan_over_kilom} in one of the channels.
The operators \(d_{\pm}\) annihilate a photon at one of the detectors and the operators \({a}_L\) and \({b}_L\) are annihilation operators of the loss channels.
Losses in the memory channel are treated analogously and we denote the efficiency up to detection (after release during a swap) as \(\eta_m\).

\deleted[id=BT]{We present a general detection model in Ref.~\cite{ART} and f}
For simplicity \replaced[id=BT]{we}{here} consider a sufficiently narrow temporal resolution and no dead-time
such that we take the temporal modes as constant within the detection window and the detectors as photon-number-resolving.
Thus, heralding on a single click ideally corresponds to post-selecting on only one detector channel \(\pm\) containing a single photon within the detector resolution \(T\) around the click time \(t_c\), i.e., a projection of the detector channels onto \(\int_{t_c-T/2}^{t_c+T/2} dt d_{\pm}^{\dag}(t) \ket{\emptyset_+} \ket{\emptyset_-}\) with the annihilation operator $d_\pm$ and vacuum state \(\ket{\emptyset_{\pm}}\) of detector channel \(\pm\).
Additionally, we model dark counts as the detection of a click event despite the corresponding detection channel being in vacuum.

While variances of the phase can be accounted for within our model by making the temporal modes complex functions, we assume phase stability or active stabilization of the setup (required by all single-click protocols \cite{sangouard-2011-quant_repeat_based_atomic_ensem_linear_optic,beukers-2024-remot_entan_protoc_station_qubit,ART}) and \replaced[id=AS]{take}{use} the modes as positive functions in the following.
Additionally, we calulate the EN \deleted[id=BT]{and BB} state perturbatively to first order in emission probabilities and the ratio of dark count to successful detections.
Within the EN links, we use the multi-mode nature of the SPDC photon pairs before the heralding to match the photon flux to the ion emission, as illustrated in Fig.~\ref{fig:SPDC-correlation}.
Slowly modulating the drive strength of the SPDC modulates the photon flux (corresponding to the slow envelopes \(\mu(t)\) of the time-bins).
This allows matching \added[id=AS]{it to} the temporal mode of the ion emission and thereby overcoming the challenge of matching a broadband ensemble system to the ion.
Because the SPDCs within the ENs match the longer ion emission time-scale, we consider them 
to be weakly driven and thus to emit uncorrelated pairs \cite{SM}.
\added[id=BT]{
  Furthermore, in this regime the slow modulation can be understood analogous to adiabatic modulation, where the operation of the SPDC source is quasi continuous.
}
The broadband nature of the SPDC emission then leads to a temporally narrow photon stored in the memory upon detection of the heralding click.
The click time is used to filter and or precisely time the release of photons from neighboring memories for entanglement swapping \deleted[id=BT]{[changed citations]}\cite{rakonjac-2021-entan_between_telec_photon_deman,askarani-2021-long_lived_solid_state_optic,teller-2025-solid_state_tempor_multip_quant}.
This makes the heralded state insensitive to \replaced[id=AS]{SPDC}{memory} photons outside the acceptance interval and
\added[id=AS]{allows us to }increas\replaced[id=AS]{e}{ing} the drive strength \replaced[id=AS]{to suppress}{thereby suppresses} part of the effect of losses in the SPDC channel.
We therefore propose to place the frequency conversion, required to match the wavelengths of the systems, in the SPDC channel.
To this end we express \(\eta' = \eta_0' \eta_{\text{FC}}\) with an intrinsic efficiency \(\eta_0'\) \added[id=BT]{(accounting for intrinsic losses due to, e.g., fiber coupling and detection)} and the frequency conversion efficiency \(\eta_{\text{FC}}\).

The state heralded by a single click in port \(\pm\) during the ion\replaced[id=AS]{-}{ }photon duration after tracing out the photon loss and detection channels is
\begin{align}
  \label{eq:RHO_EN}
  \rho_{\pm}^{\text{EN}} 
                             \approx{}& \left( A_0 \proj{0} + A_1' \proj{1} \right) \proj{\emptyset} \\
  & + A_1 \proj{\varphi_{\pm}} + A_2 \proj{1} \proj{t_c} , \notag
\end{align}
with the generated target state \(\ket{\varphi_{\pm}} = \cos\theta \ket{1} \ket{\emptyset} \pm \sin\theta \ket{0} \ket{t_c}\) with \(\tan^2\theta = \frac{{\eta' \eta_m} \abssq{\mu(t_c) \alpha_0 \beta_1}}{{\eta} \abssq{\nu(t_c) \alpha_1 \beta_0}}\).
\added[id=BT]{To simplify the target state, we used our assumption of positive temporal modes and positive amplitudes $\alpha_k$ and $\beta_k$ ($k=0,1$), different phases will lead to a phase between the logical states in $\ket{\varphi_\pm}$.}
We highlight that matching the slowly varying SPDC temporal mode \added[id=AS]{$\mu(t)$} \added[id=BT]{(for all time-bins)} to the temporal mode \added[id=AS]{$\nu(t)$} of the ion photon\added[id=AS]{, i.e. $\mu(t)\propto\nu(t)$,} makes \(\theta\) independent of the click time.
The first ket \(\ket{k}\) (\(k=0,1\)) denote the atomic state and the second ket the memory photon state in vacuum  \(\ket{\emptyset}\) or with a single excitation stored in the multi-mode memory \(\ket{t_c} = \int_{\mathbb{R}} dt F(t_c,t) c^{\dag} \ket{\emptyset}\) conditioned on the detection of a click at time \(t_c\).
\deleted[id=BT]{The elements of the density matrix up to mixed first order are given in Ref.~\mbox{\cite{ART}} and only}
\added[id=BT]{
  To leading order in the emission probabilities the coefficients in the density matrix are given by
}
\(A_0 \approx (1-\eta_m)  \frac{\tan^2\theta}{\eta_m + \tan^2\theta}\) and \(A_1 \approx \eta_m \frac{1 + \tan^2\theta}{\eta_m + \tan^2\theta}\)\replaced[id=BT]{, for the terms with}{have} a non-perturbative contribution in magnitude\replaced[id=BT]{ and }{.}
\added[id=BT]{
  the remaining elements have a linear leading order contribution
  \(A_2/\eta_m = A_1'/(1-\eta_m) = (1-\eta) \abssq{\alpha_1} \frac{\tan^2\theta}{\eta_m + \tan^2\theta}\).
}
\added[id=BT]{
  The EN ion-memory state \eqref{eq:RHO_EN} corresponds to a perfect absence-presence entangled matter-photon Bell state in the limit of perfect memory efficiency \(\eta_m=1\), low emission probability \(\abssq{\alpha_1} \to 0^+\), and balanced matching \(\tan^2 \theta = 1\).
}
\added[id=AS]{
  A full derivation including higher order contributions and dark counts, which are used in simulations, is presented in Ref.~\cite{ART}.
}

The success probability corresponding to the probability of only detecting a single click at any time \(t_c\) during the ion-pulse duration \(T_a\) in either detector \(\pm\) has the leading contribution
\begin{align}
  \label{eq:PEN}
  \mathbf{P}_{\text{EN}}
  \approx{}& \eta \abssq{\alpha_1} 
     \left( 1 + \frac{\tan^2\theta}{\eta_m} \right), 
\end{align}
if the slowly varying SPDC envelope ideally matches the ion emission \(\abssq{\mu(t_c)} = \abssq{\nu(t_c)} N\) \added[id=BT]{(for the time-bin with support containing $t_c$)}.
For simplicity, we assume that the \added[id=AS]{ion-pulse duration is divided into} \(N\) time-bins  all \replaced[id=BT]{having}{have} the same emission probability \(\abssq{\beta_1}\).
\added[id=AS]{This is done to ensure that Eq.~\eqref{eq:ini-spdc} is applicable but has no physical meaning.}
\deleted[id=AS]{In addition to the leading order, we also account for the effect of dark counts and second order terms in the emission probability in the simulations.}

\added[id=BT]{
  \emph{Entanglement Generation}---Having discussed the hybrid nodes we now want to investigate their performance within a full protocol.
  We focus on long-distance entanglement generation and thus consider a single-click scheme, but note that for shorter links or efficient ensemble memories, two-click \cite{barrett-2005-effic_high_fidel_quant_comput,sangouard-2011-quant_repeat_based_atomic_ensem_linear_optic,beukers-2024-remot_entan_protoc_station_qubit} or reflection-based \cite{nemoto-2014-photon_archit_scalab_quant_infor_proces_diamon,omlor-2025-entan_gener_using_singl_photon} protocols might be beneficial.
The entanglement generation protocol is illustrated in Fig.~\ref{fig:atom-atom-full}(d) and (e), and can be summarized in four steps.
It proceeds by (i) simultaneously generating entangle\replaced[id=AS]{d}{ment} links using single-click heralding between ions and memories  locally at ENs [Fig.~\ref{fig:atom-atom-full}(a)]
and between BB memories over a long distance \(L/n\) (with the swap level $n-1 = 1$) [Fig.~\ref{fig:atom-atom-full}(b)].
(ii) As soon as neighboring BB and EN links are ready, entanglement is swapped between them by heralding on a single click after releasing the photons from the memories [Fig.~\ref{fig:atom-atom-full}(c)].
Finally (iii) another optical swap between memories 
at the central repeater, connects the two halves.
}

\added[id=BT]{
We include a repeater in the center of the fundamental link ($n=2$), because parallelization
allows the additional swap to extend the distance, without significantly decreasing the success rate \footnote{
Further probabilistic repeaters would not benefit in the same way from the parallelization, which is why we propose to only use the probabilistic swaps in \enquote{fundamental} links and to extend range and complexity of the network beyond the fundamental link using deterministic ion swaps.
}.
Alternatively, without the central repeater \added[id=BT]{($n=1$)}, a final swap can be performed between the BB and the remaining EN.
Upon successfully heralding every step, an ion-ion entangled link is achieved.
}

\added[id=BT]{
The \emph{last part} (iv) is purification to distill two ion-ion links into a single link with enhanced fidelity, see Fig.~\ref{fig:atom-atom-full}(e).
It proceeds by preparing two ion-ion links using the steps above and performing a CNOT gate between the local ions.
Measurement of both target ions in state \(\ket{1}\) heralds the purification of the ion-ion entanglement, see Fig.~\ref{fig:atom-atom-full}(e).
This post-selection helps in suppressing events where  the memories did not store any photons while retaining the better scaling of single-click protocols with the (long-distance) fiber losses \cite{duan-2001-long_distan_quant_commun_with,simon-2007-quant_repeat_with_photon_pair,zhao-2007-robus_creat_entan_between_remot_memor_qubit,sangouard-2011-quant_repeat_based_atomic_ensem_linear_optic,beukers-2024-remot_entan_protoc_station_qubit}.
Although the single-click protocol inherently relies on phase stability \cite{sangouard-2011-quant_repeat_based_atomic_ensem_linear_optic,beukers-2024-remot_entan_protoc_station_qubit,ART},
the purification  can reduce the sensitivity to such fluctuations if the phase is stable between two successful entanglement generation events. 
}

\replaced[id=AS]{The entanglement generation between the ions and the memories have already been described above.
The simultaneous multiplexed entanglement generation in the BB has been described in multiple works \cite{duan-2001-long_distan_quant_commun_with,simon-2007-quant_repeat_with_photon_pair} and we provide an analogous model for it in Ref.~\cite{ART}.}
{In parallel to the link generation with the ENs, a multiplexed single-click protocol is performed in the BB.
These types of protocols are well established \mbox{\cite{duan-2001-long_distan_quant_commun_with,simon-2007-quant_repeat_with_photon_pair}}, and we provide a model analogous to the ENs for the BB in Ref.~\mbox{\cite{ART}}.}
\added[id=BT]{
In short, the SPDC+memory nodes in the BB can be modelled using a state of the form of Eq.~\eqref{eq:ini-spdc}, but
}
\replaced[id=BT]{d}{D}ue to the different constraints of the link generation between BB and EN, the SPDC emission probabilities within EN ($\abssq{\beta_1}$) and BB ($\abssq{\beta_1} \to \abssq{\gamma_1}$) are chosen independently.

\added[id=BT]{After entanglement generation in the EN and BB links, the generated entanglement can be swapped by reading out neighboring memories, see Fig.~\ref{fig:atom-atom-full}(c).}
Ideally, the click times and memory storage are used to re-emit the stored photons simultaneously for the optical swaps.
Therefore, we assume the memory photons share the same shape \(f(t) = F(0,t)\).
As we already included all losses in the generation step of our model to calculate the entanglement swaps, we can straightforwardly apply the same detection model used for the generation step.
The first swap (between EN and BB) ideally extends the state \(\ket{\varphi_{\pm}}\) to be between the ion and the next memory [see Fig.~\ref{fig:atom-atom-full}(d)].
Within the perturbative approximation considered here, we find the final density matrix of the ions 
\begin{align}
  \label{eq:final-state-single-rail}
  \rho = \left(\pm \alpha \op{0,1}{1,0} + \text{H.c.} \right) + \sum_{k,l=0,1} D_{k,l} \proj{k,l} ,
\end{align}
where \(\pm\) is the product of all the detectors that have clicked.
\deleted[id=BT]{The detailed calculation and density matrix elements \(D_{k,l}\) and \(\alpha\) can be found in Ref.~\cite{ART}.}
Ideally, the only non-zero elements are \(D_{0,1}=D_{1,0}=\alpha=1/2\),
which corresponds to the Bell states \(\sqrt{2} \ket{\Psi_{\pm}} = \ket{0,1} \pm \ket{1,0}\).
\added[id=BT]{The detailed calculation and density matrix elements \(D_{k,l}\) and \(\alpha\), as well as the analysis of}
\deleted[id=BT]{We analyze} the duration of \replaced[id=BT]{different}{the single link} entanglement generation \deleted[id=BT]{\(T_{\text{SL}}\) of the} protocol\added[id=BT]{s} \added[id=BT]{can be found} in Ref.~\cite{ART},
where we account for parallelization of the different entanglement generation steps using results from Refs.~\cite{jiang-2007-fast_robus_approac_to_long,coopmans-2022-improv_analy_bound_deliv_times,avis-2024-asymm_node_placem_fiber_based_quant_networ}.
We take the ENs time-scale \replaced[id=AS]{to be}{as} dominated by the ion-emission and the BBs by the quantum and classical communication time,
estimated by twice the light propagation time from the nodes to the heralding station.
The BB generation rate is directly enhanced by multiplexing via \(N_{\text{BB}}\) modes.
Additionally, the duration scales with the inverse of the first and second swap probability (which are limited by the memory efficiency\deleted[id=AS]{ and given in Ref.~\cite{ART}}).
\added[id=BT]{
  In summary, for a single ion-ion link before purification and using a central repeater the average duration is
  \(T_{\text{SL}} = \frac{3}{2} \frac{1}{\mathbf{P}_{S2}} \frac{1}{\mathbf{P}_{S1}} \frac{1}{R_{\text{BB}} + R_{\text{EN}}} \left( 1 + \frac{R_{\text{BB}}}{R_{\text{EN}}} + \frac{R_{\text{EN}}}{R_{\text{BB}}} \right)\)
  with the EN entanglement generation rate \(R_{\text{EN}} = \mathbf{P}_{\text{EN}} / \tau_{\text{EN}}\), the trial duration in the EN \(\tau_{\text{EN}}\),
  and the swap success probability of the EN-BB swap \(\mathbf{P}_{S1}\) and the final central swap \(\mathbf{P}_{S2}\).
  The backbone entanglement generation rate \(R_{\text{BB}}\) has an analogous form to the EN but is enhanced by the multiplexing capacity $N_{\text{BB}}$ and we assume the trial duration $L/(2 n c)$ (with swap level $n-1$) to be determined by the distance between nodes.
}
\begin{figure}
\centering
\includegraphics[width=\linewidth]{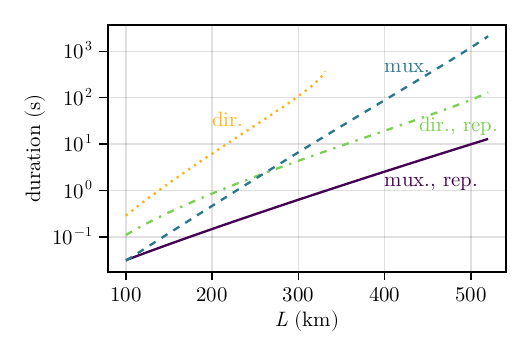}
\caption{\label{fig:duration}Duration to prepare a Bell state with \(99\%\) fidelity as a function of \replaced[id=AS]{distance for}{length comparing} different protocols.
  The dotted yellow (dash dotted green) line corresponds to direct ion-ion entanglement generation using a single-click protocol (with an ion node in the center as a repeater).
  The dashed blue (solid purple) lines correspond to the \added[id=BT]{multiplexed [mux.]} protocol proposed in this work without (with) a central multi-mode repeater \added[id=BT]{[rep.]}.
  We take the efficiencies \(\eta_m = 0.8\), \(\eta = \eta_0' = \eta_{\text{FC}} = 0.9\), and \(\eta' = \eta_0' \eta_{\text{FC}}\).
  The BB has \(N_{\text{BB}} = 1000\) multiplexing modes and an efficiency \(\eta_{\text{BB}} = \eta_0' \eta_F(L/\added[id=BT]{2}n)\) where \(\eta_F(l)\) is the fiber transmission efficiency for distance $l$ and an attenuation of \(0.2\,\)dB\(/\)km and \(n = \replaced[id=BT]{2\ (1)}{4 (2)}\) with (\added[id=BT]{with}out) a repeater.
For direct ion-ion generation the photon efficiency is \(\eta_{\text{FC}} \eta \eta_F(L\added[id=BT]{/2n})\).
Furthermore we use a dark-count rate of \(10^{-3}\,\)Hz, a pulse duration of the ions of \(10\,\){\textmu}s, a\added[id=AS]{n} acceptance window in the backbone of \(1\,\){\textmu}s (i.e., \(N=10\)), \added[id=AS]{an SPDC} correlation \replaced[id=AS]{time}{duration} of \(100\,\)ns, a detector resolution of \(1\,\)ns,
and the speed of light in fiber as \(2/3\) the vacuum speed of light.
}
\end{figure}

The fundamental link generation \added[id=AS]{protocol} corresponds to a single-click protocol, which is known to suffer from "vacuum growth" where the memories end up in the joint vacuum state \cite{sangouard-2011-quant_repeat_based_atomic_ensem_linear_optic,beukers-2024-remot_entan_protoc_station_qubit}.
This aligns with our model where in contrast to the \enquote{two-photon} component \(D_{1,1}\),
the \enquote{vacuum} component \(D_{0,0}\) always appears with a significant probability in the presence of photon loss.
\replaced[id=EH]{The inclusion of a purification step is thus necessary to limit the \enquote{vacuum} component. As discussed above,} {Therefore,} we propose to follow the generation of two fundamental links, by local CNOTs between the ions of two fundamental links.
Post-selecting on the controlled ions each being in state \(\ket{1}\) heralds success
with probability \(P_P = 2 D_{1,0} D_{0,1} + 2 D_{0,0} D_{1,1}\)
and the state of the remaining pair takes the same form as Eq.~\eqref{eq:final-state-single-rail}.
However, purification changes the \replaced[id=BT]{matrix elements}{state} according to
\(\alpha_{\text{pur}} = \frac{\alpha^2}{P_P}\) and \(D_{k,l}^{\text{pur}} = \frac{D_{k,l} D_{1-k,1-l}}{P_P}\),
so that both \(D_{0,0}^{\text{pur}}\) and \(D_{1,1}^{\text{pur}}\) can be reduced by lowering the emission probabilities \(\abssq{\alpha_1}, \abssq{\beta_1}, \abssq{\gamma_1}\) if the dark-count rate $p_d/T$ is sufficiently small, since the leading contributions of $D_{1,1}$ are proportional to these.
Assuming that the two fundamental links are generated sequentially, the complete average duration including purification is
  \(2 T_{\text{SL}} / P_P\).
  
\added[id=BT]{\emph{Performance comparison}---}With the above model, we evaluate the average duration for the entanglement of ion-ion pairs.
\replaced[id=AS]{We consider}{For} a target Bell state fidelity of \(F = 0.99\) \deleted[id=AS]{for \replaced[id=BT]{intrinsic}{internal} efficiencies of \(\eta_I = 0.9\)} and assum\replaced[id=AS]{e}{ing} the pulses to be matched, constant and of realistic durations \({1}/{\abssq{\nu(t_c)}} = {10}/{\abssq{\mu(t_c)}} = 10\,\){\textmu}s \cite{meraner-2020-indis_photon_from_trapp_ion,krutyanskiy-2023-entan_trapp_ion_qubit_separ_by_meter}.
This corresponds to matching the ion with 10 time-bins and having a time-bin duration of \(10 T_c\) \cite{lago-rivera-2021-telec_heral_entan_between_multim}.
In terms of the \replaced[id=BT]{intrinsic}{internal} efficiencies we use \(\eta\deleted[id=BT]{_A} = \eta_I\added[id=BT]{ = 0.9}\),
\(\eta\replaced[id=BT]{'}{_B} = \eta_I\added[id=BT]{^2}\) (intrinsic SPDC and frequency conversion), and \(\eta_{\text{BB}} = \eta_I \replaced[id=BT]{\eta_F(L/2n)}{e^{-L/n L_{\text{att}}}}\) \added[id=BT]{with the fiber transmission efficiency $\eta_F(l) = e^{-l/ L_{\text{att}}}$, the swap level} \(n-1 = \replaced[id=BT]{1\ (0)}{4 (2)}\)\replaced[id=BT]{, and the fiber attenuation length}{with(out) a repeater and} \(L_{\text{att}}=\frac{10}{0.2 \log 10}\,\text{km}\).
\added[id=BT]{We use a memory efficiency of \(\eta_m = 0.8\).}
Optimizing with respect to the emission probabilities \(\abssq{\alpha_1}, \abssq{\beta_1}, \abssq{\gamma_1}\), see Refs.~\cite{SM,ART,CODE}, we find the results shown in Fig.~\ref{fig:duration}. 

For comparison, we also include the time for direct ion-ion entanglement.
Due to the longer ion pulses, direct generation only works for a shorter distance for the \replaced[id=BT]{considered}{same} dark-count rate.
We observe a speedup of around an order of magnitude thanks to the multiplexing of the hybrid architecture compared to direct ion-ion links.
While this does not saturate the ideal speedup of the channel capacity due to the additional probabilistic steps and intrinsic losses, it provides a significant speed-up compared to the \replaced[id=BT]{direct}{BB less} case.
For all protocols included in the figure, the entangled state is between ions, and thus the advanced trapped-ion capabilities are available \added[id=AS]{for further processing}, including deterministic entanglement swaps\deleted[id=AS]{ to extend the range}.
\added[id=AS]{In the absence of dark counts the duration is proportional to $1/(1-F)$ to leading order.} If the target fidelity is relaxed to \(F=0.9\), \replaced[id=BT]{the proposed}{our} protocol including a central multi-mode memory repeater achieves an average entanglement generation rate exceeding \(1\,\)Hz over a distance of \(500\,\)km \replaced[id=AS]{compared to \(0.1\,\)Hz for $F=0.99$}{(for the same parameters)}.

\replaced[id=AS]{W}{Finally, w}hile we here considered a single-click protocol with purification of the ions, we show in Ref.~\cite{ART}, for high memory efficiencies, we can achieve similar rates between using a double-click approach to create spin-photon entanglement between the ion and a dual rail setup. This relaxes the phase stability requirements on the ions and omits the need of an extra purification step.
Therefore, we consider this approach more viable in the presence of highly efficient photonic memories, whereas the protocol investigated in this work is more suitable for near-term memories.

\emph{Conclusion}
We developed a protocol uniting trapped ions with atomic ensemble systems.
At the core of our protocol is a novel approach to link spontaneous parametric down conversion sources and absorptive memories with trapped ions, i.e., broad-band ensemble systems with single narrow-band quantum emitters.
This enables the use of multiplexing of ensemble based approached combined with the more advanced gate sets and deterministic entanglement swaps of trapped ions for further processing of the information.
Our approach shows a significant speedup and provides better resilience regarding dark counts compared to direct ion-ion entanglement generation.
Moreover, the SPDC matching approach can be extended to couple other narrow‑band systems to broadband memories, thereby also enabling inter-connectivity between those systems.

\begin{acknowledgments}
  We thank T. E. Northup, H. de Riedmatten, S. D. C. Wehner, M. van Hooft, N. Sangouard, P. Cussenot, B. Grivet, S. Grandi, \deleted[id=BT]{and} A. Das\added[id=BT]{, and C. Gustin} for fruitful discussions.
    This work was funded by the European Union's Horizon Europe research and innovation programme under grant agreement No.~101102140 – QIA Phase 1.
  Funded by the European Union.
  Views and opinions expressed are however those of the authors only and do not necessarily reflect those of the European Union or European Commission.
  Neither the European Union nor the granting authority can be held responsible for them.
  BT, ERH and ASS acknowledge the support of Danmarks Grundforskningsfond (DNRF Grant No.~139, Hy-Q Center for Hybrid Quantum Networks).
\end{acknowledgments}

\bibliography{refs.bib}

\appendix
\clearpage
\section{SPDC uncorrelated pairs}\label{app:uncor}
To match the SPDC and trapped ions, we considered the weak driving limit for the SPDC in the main text.
In this appendix we quantify the meaning of weak driving, link it to the emission of uncorrelated photon pairs and the implications for the model used in the main text.
Considering a weakly driven SPDC source, it is unlikely that two emissions happen simultaneously.
Here, we quantify the meaning of ``weakly driven'' as having a small probability to have a photon pair emitted during time-bin duration \(T_{\text{TB}} \gg T_c\) (with the correlation time $T_c$).
Thus we write the two-photon component using \(G(\vec{t}) = \delta_1 F(t_1,t_3) F(t_2,t_4) + \delta_2 (t_1 \leftrightarrow t_2)\).
Normalization of the two-photon wave function leads to \(2 \abssq{\delta_1 + \delta_2} \approx 1\), where we disregarded terms of \(\mathcal{O}(T_c/T_{\text{TB}}\)).

Additionally, we assume that in the weakly driven case, the photons in a single channel should be uncorrelated (or the emitted photon pairs should be uncorrelated).
This corresponds to a \(g_2(t,t') \approx 1\) within channel \(b\) or to
\begin{align}
  \label{eq:correlation-relation}
  \braket{ \Psi_b | b^{\dag}(t)b(t) | \Psi_b } \approx \frac{\braket{ \Psi_b | b^{\dag}(t) b^{\dag}(t') b(t') b(t) | \Psi_b }}{\braket{ \Psi_b | b^{\dag}(t') b(t') | \Psi_b }},
\end{align}
for \(\abs{t-t'} \gg T_c\).
The vacuum part of the state [i.e., the term \(\propto \beta_0\) in Eq.~(2) \added[id=BT]{of the main text}] does not contribute to either side of Eq.~\eqref{eq:correlation-relation}.
Evaluating the expectation values of Eq.~\eqref{eq:correlation-relation} leads to
\begin{align}
  \label{eq:correlation-condition}
  \abssq{\beta_1} + 2 \abssq{\beta_2} = \frac{2 \abssq{\beta_2}}{\abssq{\beta_1} + 2 \abssq{\beta_2}} ,
\end{align}
which is solved by \(2 \abssq{\beta_{2,\pm}} = \frac{1}{2} - \abssq{\beta_1} \pm \sqrt{\frac{1}{4} - \abssq{\beta_1}} \approx \begin{cases} 1 - 2 \abssq{\beta_1} - \abs{\beta_1}^4 \\ \abs{\beta_1}^4 \end{cases}\).
The approximation holds for small \(\beta_1\) and we expect \(\abssq{\beta_2} < \abssq{\beta_1}\),
therefore we use \(2 \abssq{\beta_2} \approx \abs{\beta_1}^4\) for small emission probabilities (or weak drives).

\section{\added{Derivation of the Hybrid Edge-Node-ion-memory-photon Entangled State}}\label{app:uncor}
\added[id=BT]{
In the main text we presented the main idea of the EN entanglement generation between the ion and a photon stored within the multimode memory, and in a companion article~\cite{ART} we provide a detailed derivation for states and probabilities involved in the full protocol as well as a comparison between double and single click protocols.
For completeness we summarize the key steps of the derivation to calculate the EN density matrix here.
}

\added[id=BT]{
  We start from the product state of the two subsystems \(\ket{\Psi_a} \ket{\Psi_b}\) with the individual states defined according to Eqs.~(1) and (2) of the main text.
  Note that for the SPDC+M state we only account for the time-bin containing the heralding click.
  Furthermore, we assume that frequency conversion is used to match the carrier frequencies and the systems are driven such that SPDC and ion emission share the same envelope.
  The which path information is removed using a beamsplitter and the success of the attempt is heralded by a single click.
The non-normalized state describing the success probability and resulting state without dark-counts is thus given by
}
\begin{widetext}
\begin{align}
  \label{eq:click_state_EN}
  \ket{\Psi}_{\text{EN},1} = \int_{\mathcal{T}} dt & \bra{\emptyset}_d d_{\pm}(t) \ket{\Psi_a} \ket{\Psi_b} \\
  \approx
 \int_{\mathcal{T}} dt & \Big\{
\alpha_1 \left[ \beta_0 + \beta_1 \sqrt{1-\eta'} \int_{\mathbb{R}^2} dt'dt'' \mu(t') {b}_L^{\dag}(t') F(t',t'') c(t') \right] \ket{1} \sqrt{\frac{\eta}{2}} \nu(t) \notag \\
& \ \pm \alpha_0 \left[ \beta_1 + \beta_1^2 \sqrt{1-\eta'} \int_{\mathbb{R}^2} dt'dt'' \mu(t') {b}_L^{\dag}(t') F(t',t'') c^{\dag}(t') \right] \ket{0} \sqrt{\frac{\eta'}{2}} \mu(t) \int_{\mathbb{R}} d\tau F(t,\tau) c^{\dag}(\tau) \notag \\
& \  + \alpha_1 \beta_1 \ket{1} \int_{\mathbb{R}} dt' \nu(t') \sqrt{1-\eta} {a}_L^{\dag}(t) \sqrt{\frac{\eta'}{2}} \mu(t) \int_{\mathbb{R}}d\tau F(t,\tau) c^{\dag}(\tau) \Big\} \ket{\emptyset} , \notag
\end{align}
\end{widetext}
\added[id=BT]{
  where we still include the loss channels in the state and
  denote the vacuum state in both detection channels as \(\ket{\emptyset}_d\).
  The remaining symbols are defined in the main text.
Furthermore, we used the uncorrelated pair emission discussed in the previous section of the supplemental material to reduce the amount of parameters.
}

\added[id=BT]{
Dark counts are modeled as observation of a click while the channel is in vacuum.
Within the perturbative analysis we use, we treat the dark count probability \(p_d\) in the time-window of duration \(T\) to be of comparable order to the mixed second order of the emission probabilities \(\abssq{\alpha_1}\) and \(\abssq{\beta_1}\) (note that ideally it should be smaller than that to have a decent fidelity), such that the relevant non-normalized state heralded by a dark count reads
\begin{align}
  \label{eq:dark_count_state_EN}
  \ket{\Psi}_{\text{EN},0} = \sqrt{p_d} \bra{\emptyset}_d \ket{\Psi_a} \ket{\Psi_b} \approx \sqrt{p_d} \ket{0} \ket{\emptyset},
\end{align}
where \(\ket{\emptyset}\) is the vacuum state in all the remaining optical channels, including the loss channels and the memory channel.
}

\added[id=BT]{
The non-normalized density matrix is then given by first applying the memory loss model \(c \to \sqrt{\eta_m} c + \sqrt{1-\eta_m} c_L\) and then tracing out all photonic loss modes on the sum of the non-normalized states in Eqs.~\eqref{eq:click_state_EN} and Eq.~\eqref{eq:dark_count_state_EN}, i.e. \(P_{\pm} \rho_{\pm}^{\text{EN}} = \tr_{\text{loss}} \sum_{k=0,1}\proj{\Psi}_{\text{EN},k}\).
Justified by the temporal filtering and the uncorrelated SPDC emission, we additionally treat the un-conditioned memory photons [the $F(t',t'')c^{\dag}(t')$ terms in Eq.~\eqref{eq:click_state_EN}] as an extra channel that we trace out.
From this the density matrix given in the main text in Eq.~(3) is calculated.}

\added[id=BT]{
For simplicity we focused on the most relevant time-bin in the preceding section, the full SPDC+M state, however, is spanned by multiple time-bins to match the full ion emission, i.e., $\sum_{i=1}^N \mu_i(t) = N \nu(t)$ with the index $i$ labeling the time-bin.
To account for the SPDC+M nodes state using \(N\) time-bins, we need to ensure that no click was observed from the remaining \(N-1\) time-bins.
If each time-bin has the same emission probability $\abssq{\beta_1}$, the probability to be in vacuum to first order in the emission probability is \(p \approx 1 - \eta' \abssq{\beta_1}\) or for all \(N-1\) time-bins \(p^{N-1} \approx 1 - (N-1) \eta'\).
The total success probability per attempt in the main text is given by integrating the click rate over all possible click times during the ion emission (\(\mathbb{T}_I\)) and summing over the possible detector outcomes; this leads to \(\mathbf{P}_{\pm} = p^{N-1} \int_{\mathbb{T}_I} dt_c  \frac{P_+ + P_-}{T}\).
}

\section{Additional details on the optimization}\label{app:opt} 

\begin{figure}[ht]
\centering
\includegraphics[width=8cm]{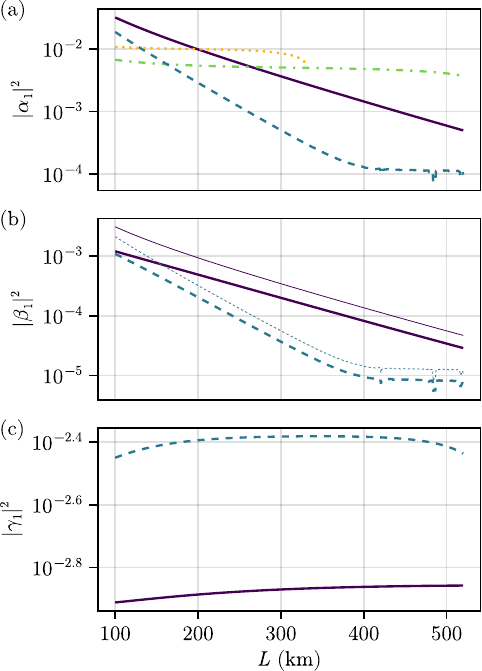}
\caption{\label{fig:optimal_emission_probabilities}Optimal emission probabilities corresponding to the results displayed in Fig.~3 of the main text.
  The line styles encode the protocols and follow Fig.~3 of the main text.
  The thin lines in (b) correspond to the semi-analytic approach [see Eq.~\eqref{eq:opt_theta}].
  In panel (a) we show the emission probability of the ions \(\abssq{\alpha_1}\), in (b) of the SPDC connecting memory and ions in the end-nodes \(\abssq{\beta_1}\) and in (c) of the SPDC within the BB \(\abssq{\gamma_1}\).}
\end{figure}

The emission probabilities corresponding to the results in Fig.~3 of the main text are displayed in Fig.~\ref{fig:optimal_emission_probabilities}.
For the protocol proposed in the main text, we numerically optimize the emission probabilities
\(\abssq{\alpha_1}, \abssq{\beta_1},\) and \(\abssq{\gamma_1}\)
using \enquote{Optim.jl} \cite{mogensen-2018-julia_optim}.
For the direct ion-ion approach, we use \(10000\) exponentially spaced samples for the ion emission probability \(\abssq{\alpha_1}\) between \(10^{-1}\) and \(10^{-6}\), decreasing until we find the first value that satisfies the fidelity constraint,
while for the proposed protocol, we perform a numerical optimization of all three emission probabilities.
We provide the implementation in Ref.~\cite{CODE}.

To further understand the emission probabilities,
we estimate the optimal asymmetry in the generated link at the edge-nodes, quantified by the angle \(\theta\).
To this end, we analytically maximize the leading-order product \(\mathbf{P}_{S1} \mathbf{P}_{S2} \alpha\), which is 0th order in emission amplitudes.
For completeness we note that the leading orders success probabilities of the first and second swap operations are
\(\mathbf{P}_{S1} \approx \frac{\eta_m}{2} + \eta_m (1-\eta_m) \frac{\tan^2\theta}{\eta_m + \tan^2\theta}\) and
\(\mathbf{P}_{S2} \approx \frac{(3+X) (1-\eta_m) \sin^2\theta + 2 \eta_m \cos^2\theta}{\eta_m + 3 (1-\eta_m) \sin^2\theta} \frac{\eta_m \sin^2\theta}{\eta_m + X (1-\eta_m) \sin^2\theta}\),
where \(X=1(3)\) without (with) the central repeater.
For more details see Ref.~\cite{ART}.
Finally, for the leading order off-diagonal density matrix element we have
\(\alpha \propto \frac{1}{2} \cos^2\theta \sin^2\theta \frac{\eta_m}{\eta_m + 3(1-\eta_m)\sin^2\theta} \frac{\eta_m}{\eta_m + X(1-\eta_m)\sin^2\theta}\).

Analytically we find a maximum of \(\mathbf{P}_{S1} \mathbf{P}_{S2} \alpha\) with regard to \(\theta\) if \(\theta\) satisfies
\begin{align}
  \label{eq:opt_theta}
  \tan^2 \theta = \frac{\eta_m}{\sqrt{\eta_m + (1 - \eta_m) X}} .
\end{align}
Note that the (ideal) final state remains symmetric, because the asymmetry of the EN states contributes evenly on both sides (i.e., symmetric).
A comparison of the emission probabilities predicted using the semi-analytical approach and the numeric approach is given in Fig.~\ref{fig:optimal_emission_probabilities}.

\end{document}